\documentclass[preprint,showpacs,preprintnumbers,amsmath,amssymb,floatfix]{revtex4}

\usepackage{graphicx}
\usepackage{dcolumn}
\usepackage{bm}

\newcommand{\Vp}{{\bm p}}

\makeatletter
\def\lsim{\mathrel{\mathpalette\gl@align<}}
\def\gsim{\mathrel{\mathpalette\gl@align>}}
\def\gl@align#1#2{\lower.6ex\vbox{\baselineskip\z@skip\lineskip.3ex
     \ialign{$\m@th#1\hfill##\hfil$\crcr#2\crcr\sim\crcr}}}
\makeatother

\begin{document}

\preprint{MAP-335}


\title{From low-density neutron matter to the unitary limit}

\author{T. Abe${}^{1}$ 
\footnote{Current address: Center for Nuclear Study, Graduate School of Science, the University of Tokyo,
RIKEN Campus, Wako, Saitama 351-0198, Japan}
and R. Seki${}^{2}$}

\affiliation{
${}^1$
Department of Physics, Tokyo Institute of Technology,
Megro, Tokyo 152-8551, Japan\\
${}^2$ Department of Physics and Astronomy, California State University,
Northridge, Northridge, CA 91330, USA
}

\date{\today}

\begin{abstract}

Various quantities of an attractively interacting fermion system
at the unitary limit are determined by extrapolating Monte Carlo
results of low-density neutron matter.  Smooth extrapolation in
terms of $1/(k_F a_0)$ ($k_F$ is the Fermi momentum, and $a_0$ is
the ${}^1S_0$ scattering length) is found with the quantities
examined: the ground-state energy, the pairing gap at $T
\approx 0$, and the critical temperature of the
normal-to-superfluid phase transition. We emphasize proximity of
the physics of low-density neutron matter to that at the unitary
limit. The extrapolated quantities are in a reasonable agreement
with those in the literature.

\end{abstract}
\maketitle

\section{Introduction}
\label{Introduction}

In our previous paper \cite{abeseki}, we reported a Monte Carlo
calculation of thermodynamic properties of low-density neutron
matter by using nuclear effective field theory (EFT) \cite{eft}.
As pairing in neutron matter is strong, neutron matter is a
strongly correlated fermionic system.  We firmly established that
low-density neutron matter is in the state of BCS-Bose-Einstein
condensation (BEC) crossover \cite{crossover} instead of a
BCS-like state, the standard description in nuclear physics
\cite{abook}.

The crossover state of low-density neutron matter is actually an
expected one from studies of the BCS-BEC crossover over the past
decade. The pairing strength of an attractively interacting
(therefore unstable) fermion system is characterized by a product
of the two physical parameters, the Fermi momentum $k_F$ and the
($S$-wave) scattering length $a_0$ \cite{Randeria}. In terms of
$\eta \equiv 1/(k_F a_0)$, the state of the BCS limit is realized
at $\eta \rightarrow -\infty$, that of the BEC limit at $\eta
\rightarrow +\infty$, and that of the unitary limit at $\eta
\rightarrow 0$. (We use the sign convention $a_0 < 0$ for
fermions attractively interacting with no bound state.) A fermion
pair at the limit forms a zero-energy bound state, thereby
yielding an infinitely long scattering length.  Because the
infinitely long scattering length generates no classical scale,
the fermion system composed of the pairs is expected to have a
universal feature. In recent years, much attention has been paid
to the physics at the unitary limit in the fields of both
condensed matter and atomic physics; for example, 
Refs.~\cite{burovski,thermodynamics_unitary_limit}, 
and references therein. The
${}^1S_0$ neutron-neutron scattering length is negative and large
in magnitude, $a_0 \approx -16.45$ fm \cite{Noyes:1972}.
For a density of about
$(10^{-4}-10^{-2})\rho_0$ ($\rho_0 = 0.16$ fm$^{-3}$ is the nuclear matter
density), $\eta$ takes the value of
\begin{equation}
-0.8 \alt \eta \alt -0.2 . \label{etaldnm}
\end{equation}
Furthermore, the neutron matter of the above density is well
described by the EFT lattice Hamiltonian that is identical to the
attractive Hubbard model \cite{ask}. Equation~(\ref{etaldnm}) thus
suggests that the physics of low-density neutron matter would be
similar to that of the unitary limit. Note that the similarity is
expected only in the limited range of the density. For $\eta
\ll -0.8$ ($\eta \rightarrow - \infty$), neutron matter would be
in more of a BCS state in this quite low density. For $\eta \agt
-0.2$ ($\eta \rightarrow 0$), the physics of neutron matter
is much affected by the next-order (repulsive) term in the EFT
potential, and deviates from that of the unitary limit.

Physically, neutron matter never reaches the unitary limit $\eta =
0$. But our previous Monte Carlo results at the leading-order (LO)
\cite{abeseki} are based on the Hubbard model with the parameter
values chosen (translated from $k_F$ and $a_0$) to be suitable for
neutron matter.  A Monte Carlo calculation of a fermion system at
the unitary limit could be carried out by repeating our
neutron-matter calculation but with the parameter values adjusted
to the limit.  This approach has been used by Lee
\cite{Lee:2006,Lee:2005it,Lee:2008xs, Lee:2008fa}, based on the
Lagrangian formalism.  While it may appear to be a
straight-forward application, the computation is quite time
consuming. Instead, by exploiting the proximity of low-density
neutron matter to the unitary limit, we follow here another
procedure originally proposed in Ref.~\cite{ck}, an extrapolation to
the unitary limit from low-density neutron matter by use of our
previous Monte Carlo results at the LO.

The underlying assumption in this procedure is that the
physical quantities of interest are, at least numerically, slowly
varying functions of $\eta$ and smoothly reachable to the unitary
limit, as Eq.~(\ref{etaldnm}) suggests. As will be seen in
Secs.~III and IV, we find that this is indeed the case. The
quantities examined are the ground-state energy $E_{g.s.}$, the
pairing gap at $T \approx 0$ $\Delta$, and the critical
temperature of the normal-to-superfluid phase transition $T_c$,
all of which are made to be dimensionless by taking ratios with
$\epsilon_F \equiv k_F^2/(2M)$ ($M$ is the neutron mass).  Note that
only $k_F$ provides a classical dimension at the unitary limit in
the system.

This paper is organized as follows: After the Introduction. 
Sec.~II briefly summarizes our computational method,
which will help keep the discussions in later sections
coherent. In Secs.~III, IV, and V, we describe the
extrapolation procedure of $E_{g.s.}$, $\Delta$, and $T_c$,
respectively.  The description of $E_{g.s.}$ in Sec.~III is
somewhat detailed, since this quantity was not examined in our
previous paper \cite{abeseki}. Our conclusion is found in Sec.~VI.

\section{Monte Carlo Computation for the Standard Parameter Set}

The neutron-neutron ($nn$) interaction in the EFT Lagrangian
includes all possible terms allowed by symmetries of the
underlying theory of quantum chromodynamics (QCD) \cite{eft}. The
$nn$ potential is in the momentum expansion form
\begin{equation}
V({\Vp'},{\Vp}) = c_0(\Lambda) +c_2(\Lambda)({\Vp}^2+{\Vp'}^2)+
\cdots - 2c_2(\Lambda){\Vp}\cdot{\Vp'}+ \cdots, \label{EFTpot}
\end{equation}
where ${\Vp}$ and ${\Vp'}$ are the $nn$ center-of-mass momenta, and
$\Lambda$ is the regularization scale.  For a description of
low-density neutron matter, we use a truncated potential including
only the first term in Eq.~(\ref{EFTpot}).  The truncated EFT
Hamiltonian on a three-dimensional cubic lattice ${\hat H}$
then takes the form of the three-dimensional attractive 
Hubbard-model Hamiltonian
\cite{ask},
\begin{eqnarray}
  {\hat H} &=& - t \sum_{\langle i,j \rangle \sigma}
  {\hat c}_{i \sigma}^\dagger {\hat c}_{j \sigma}
      + 6t\sum_{i\sigma}
      {\hat c}_{i\sigma}^\dagger {\hat c}_{i\sigma}
  + \frac{1}{a^3}c_0(a) \sum_{i}
  {\hat c}_{i \uparrow}^\dagger
  {\hat c}_{i \downarrow}^\dagger
  {\hat c}_{i \downarrow}
  {\hat c}_{i \uparrow}\nonumber\\
  &=& - t \sum_{\langle i,j \rangle \sigma}
      {\hat c}_{i\sigma}^\dagger {\hat c}_{j\sigma}
      + 6t \sum_{i\sigma} {\hat n}_{i\sigma}
      + \frac{1}{a^3} c_0(a) \sum_{i} {\hat n}_{i\uparrow} {\hat
      n}_{i\downarrow},
   \label{hubbard}
\end{eqnarray}
where $a$ is the lattice spacing, $t = 1/(2M a^2)$ is the hopping
parameter ($M$, the neutron mass), and $\langle i,j \rangle$
denotes a restriction on the sum of all neighboring pairs. ${\hat
c}_{i \sigma}^\dagger$ and ${\hat c}_{i \sigma}$ are the creation
and annihilation operators of the neutron, respectively ($\sigma =
\uparrow, \downarrow$), and ${\hat n}_{i\sigma} = {\hat
c}_{i\sigma}^\dagger {\hat c}_{i\sigma} $ is the number operator
with the spin $\sigma$ at the $i$ site.

The lattice spacing $a$ is directly related to $\Lambda$ as
\begin{equation}
  \Lambda \sim \frac{\pi}{a}.
\label{lambda1}
\end{equation}
$\Lambda$ should be set larger than the momentum scale, below
which the truncated form of the potential is valid \cite{bira}.
$\Lambda$ can be chosen to be smaller but should be at least
\begin{equation}
  \Lambda > p\
\label{lambda2}
\end{equation}
for the momentum $p$ at which the physics is studied
\cite{lepage}.  In our case, we take $p \sim k_F$, because the
momentum scale corresponding to the excitation energy of interest
is much less.

The EFT parameter $c_0(a)$ in the lattice regularization with the
finite lattice spacing $a$ is given as \cite{sv}
\begin{equation}
c_0(a)=\frac{4\pi}{M}\left(\frac{1}{a_0}-\frac{2\theta_1}{a}\right)^{-1},
\label{EFTpara}
\end{equation}
where
\begin{equation}
\theta_1 \equiv \frac{1}{8\pi^2}
\wp\int^{\pi}_{-\pi}\int^{\pi}_{-\pi}\int^{\pi}_{-\pi}
\frac{dx \: dy \; dz}{3-(\cos x +\cos y +\cos z)}=1.58796\cdots
\label{theta1}
\end{equation}
is a form of Watson's triple integral \cite{watson}, with $\wp$
denoting the principal value of the integral. At the unitary limit
($|a_0| \rightarrow \infty$), we have
\begin{equation}
c_0(a)/(a^3 t) \rightarrow - 4\pi/\theta_1 = - 7.91353\cdots,
\label{potunitary}
\end{equation}
which agrees with the value found in the literature
\cite{burovski}.

In our Monte Carlo calculation at the LO \cite{abeseki}, we
carried out an extensive Monte Carlo lattice calculation for the
standard parameter set, which consists of the three values of the
neutron matter density $\rho$ with the lattice filling (or the
site-occupation fraction) $n$ set to be 1/4. The reasoning
underlying this choice is somewhat involved, and we refer the
reader to our previous paper \cite{abeseki}.  Here, $n$ is defined
as
\begin{equation}
  n \equiv a^3\rho = \frac{1}{N_s} \sum_{i,\sigma}
  \langle {\hat c}_{i\sigma}^\dagger {\hat c}_{i\sigma} \rangle,
  \label{nrho}
\end{equation}
where $N_s$ is the lattice size (the total number of the three-dimensional 
lattice sites). Note that in this work, we specify the density of
the interacting fermion system, $\rho$, using $k_F$ through
the relation $\rho = k_F^3/(3\pi^2)$. We emphasize that $\rho$
here is an expectation value obtained by our Monte Carlo
calculation, as Eq.~(\ref{nrho}) shows.

The three densities of the standard parameter set are $k_F =$ 15,
30, and 60 MeV.  The corresponding three values of $\eta$ are
listed in Table I, together with those of $a$ and $c_0(a)$, which
follow from the $k_F$ values with $n=1/4$. Though not used in the
Monte Carlo calculation, the value of $c_0(a)$ at the unitary
limit, Eq.~(\ref{potunitary}), is also shown in the table for
comparison. Note that $a$ in the standard parameter set satisfies
the EFT regularization condition, Eq.~(\ref{lambda2}).

\begin{table}[htbp]
\caption{Standard parameter set and $c_0(a)/(a^3 t)$ at the
unitary limit $\eta =0$.} \label{table:3eta}
    \begin{center}
    \begin{tabular}{ccccc}
      \hline
      \hline
      $\eta$ & $c_0/(a^3t)$ & $k_F$ (MeV) & $a$ (fm) & $\Lambda$ (MeV) \\
      \hline
      $-0.7997$ & $-5.308$ & $15$ & $25.64$ & $24.18$ \\
      $-0.3999$ & $-6.354$ & $30$ & $12.82$ & $48.36$ \\
      $-0.1999$ & $-7.049$ & $60$ & \ \, $6.409$ & $96.73$ \\
      \ \, $0.0000$ & $-7.914$ &      &         &         \\
      \hline
      \hline
    \end{tabular}
  \end{center}
\end{table}

The Monte Carlo calculations were performed on cubic lattices of
$N_s = 4^3$, $6^3$, $8^3$, and $10^3$ by the method of
determinantal quantum Monte Carlo
\cite{LohGubernatis1992,dosSantos2003}, commonly used in
condensed-matter physics. Using the data on the four different
$N_s$'s, we apply the method of finite-size scaling to extrapolate
to the thermodynamic limit. We take the continuum limit ($n
\rightarrow 0$) by extrapolation using Monte Carlo data for
various $n$'s on the $N_s = 6^3$ lattice. In the following section
we elaborate on how we determine $E_{g.s.}$ at these limits and how
we then extrapolate to $\eta \rightarrow 0$.

\section{Ground-state energy $E_{g.s.}$}

Following common practice, we express the ground-state energy per
particle of neutron matter $E_{g.s.}$ in terms of the energy
parameter $\xi$,
\begin{equation}
  \xi = \frac{E_{g.s.}}{E_{FG}} = \frac{5 E_{g.s.}}{3 \epsilon_F},
\label{defxi}
\end{equation}
where $E_{FG}$ is the ground-state energy per particle of the
corresponding noninteracting system. $\xi$ is expected to be of a
universal character at the unitary limit. As done in our previous
paper, we determine $\xi$ at the three values of $\eta$ from the
LO lattice calculations, first by taking the thermodynamic
limit and second by applying the continuum limit. After
carrying out the two steps, we extrapolate $\xi$ to $\eta=0$.

\begin{figure}[htbp]
\begin{center}
\includegraphics[width=100mm]{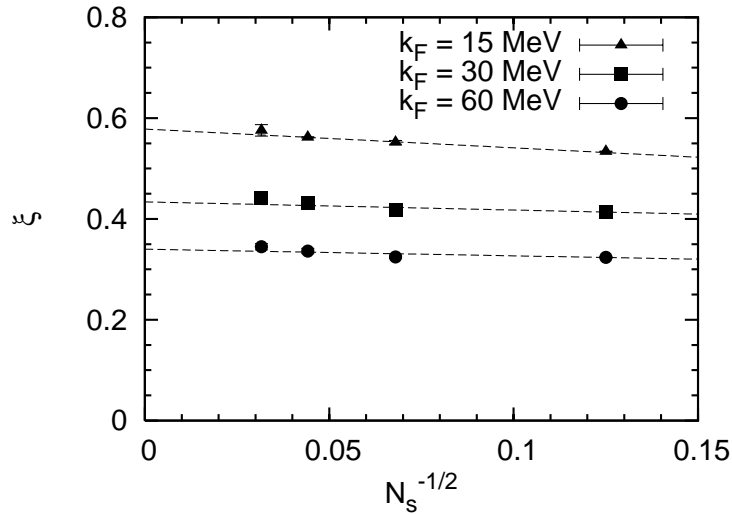}
\caption{Finite-size scaling of the energy parameter $\xi$. The
thermodynamic limit is at $N_s \rightarrow \infty$. The Monte
Carlo data of $\xi$ with statistical uncertainties are shown at
the Fermi momentum of $k_F = 15$ (triangle), $30$ (square), and
$60$ MeV (circle). The dotted lines are the $N_s^{-1/2}$ linear
fits, Eq.~(\ref{Eq:fitting_xi}), to the data of $N_s=4^3$, $6^3$,
$8^3$, and $10^3$.} \label{Fig:xi.fss}
\end{center}
\end{figure}

First, we determine $\xi$ of the thermodynamic limit at the
three $\eta$ by applying the method of finite-size scaling using
Monte Carlo data for the four lattice sizes, $N_s = 4^3$, $6^3$,
$8^3$, and $10^3$. The data used for the determination are
shown with statistical uncertainties in Fig.~\ref{Fig:xi.fss}.

As in the case of $\Delta$ discussed in Ref.~\cite{abeseki}, the
$N_s$ dependence of $\xi$ is found to be weak.  The scaling
exponent is difficult to determine, and the best fit to the Monte
Carlo data results in an exponent with a large uncertainty,
essentially being indefinite.  As shown in Fig.~\ref{Fig:xi.fss},
we find the choice of the $N_s$ scaling power of $\xi$, $\sim
N_s^{-1/2}$ (the same as that of $\Delta$ \cite{abeseki}) works
reasonably well though not quite ideally.  Because of the limited
number of our Monte Carlo data, we decided to proceed with the
analysis using the $N_s^{-1/2}$ scaling.  With this scaling, the
best fits are found to be
\begin{eqnarray}
\label{Eq:fitting_xi}
  \xi (\eta \approx -0.8, N_s)
  &=& -0.37(13) \ N_s^{-1/2} + 0.5784(35), \nonumber \\
  \xi (\eta \approx -0.4, N_s)
  &=& -0.162(84) \ N_s^{-1/2} + 0.4339(75), \\
  \xi (\eta \approx -0.2, N_s)
  &=& -0.131(63) \ N_s^{-1/2} + 0.3400(80), \nonumber
\label{xivalues}
\end{eqnarray}
and are shown with the Monte Carlo data in Fig.~\ref{Fig:xi.fss}.
The last constant in each equation in Eq.~(\ref{Eq:fitting_xi}) is
$\xi$ in the thermodynamic limit ($N_s \rightarrow \infty$).

The preceding best fits are performed using the jackknife method
(often used in the lattice QCD data analysis \cite{LatticeQCD}).
The method is used to obtain all best fits in this work.

\begin{figure}[htbp]
\begin{center}
\includegraphics[width=100mm]{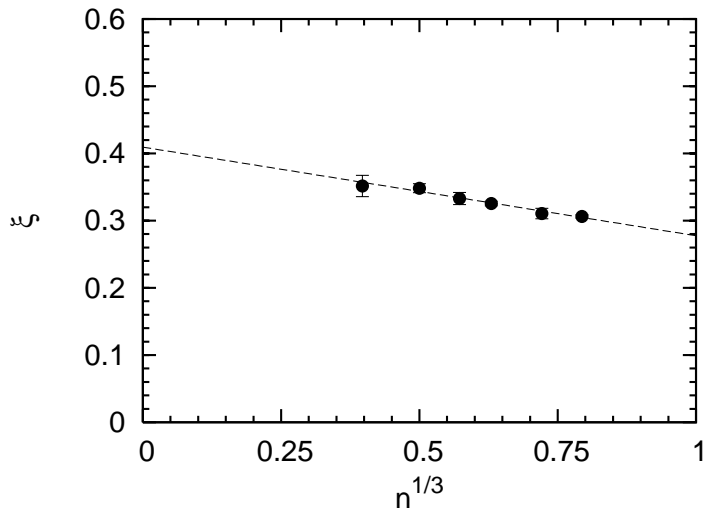}
\caption{Energy parameter $\xi$ as a function of the lattice
filling $n$.  Monte Carlo data (solid circles with
statistical uncertainties) are for $n = 1/16$, $1/8$, $3/16$,
$1/4$, $3/8$, and $1/2$ on the $N_s = 6^3$ lattice for $k_F = 60$
MeV. The dashed line, Eq.~(\ref{nfit}), is the best fit to the
data.} \label{Fig:xi_adep}
\end{center}
\end{figure}

Second, we take the continuum limit through $n \rightarrow 0$ in a
procedure similar to that in Ref.~\cite{burovski}, 
as discussed in Ref.~\cite{abeseki}.  
As shown in Fig.~\ref{Fig:xi_adep}, the Monte
Carlo data for the lattice fillings of $n = 1/16$, $1/8$, $3/16$,
$1/4$, $3/8$, and $1/2$ are best fit by
\begin{equation}
  \xi(n) = - 0.132(24) \ n^{1/3} + 0.409(15).
\label{nfit}
\end{equation}
Note that $n=1/2$ corresponds to the quarter filling of the
lattice. The data are taken on a lattice of $N_s = 6^3$ at the
Fermi momentum $k_F = 60$ MeV.  Equation~(\ref{nfit}) gives the
ratio of $\xi(n \rightarrow 0)$ and $\xi(n = 1/4)$ as 1.26(6).

As discussed in Ref.~\cite{abeseki}, such a ratio is expected and
is also confirmed to depend weakly on $N_s$ and $k_F$ in the case
of $\Delta$ at $T \approx 0$, $T_c$, and the pairing temperature
scale $T^\ast$.  Here, we assume that $\xi$ also weakly depends on
$N_s$ and $k_F$.  Applying the same ratio to $\xi$ at the
thermodynamic limit, we then obtain
\begin{eqnarray}
\label{Eq:xi_finite_final}
  \xi (\eta \approx -0.8) &=& 0.728(37), \nonumber \\
  \xi (\eta \approx -0.4) &=& 0.546(34), \\
  \xi (\eta \approx -0.2) &=& 0.428(30). \nonumber 
\end{eqnarray}
The three values of $\xi$ of Eq.~(\ref{Eq:xi_finite_final})
are best fitted by
\begin{equation}
  \xi = 0.292(24) - 0.795(33) \ \eta - 0.271(21) \ \eta^2 ,
\label{eq:xi_extrapolation}
\end{equation}
which yields $\xi= 0.292(24)$ at the unitary limit by setting
$\eta \rightarrow 0$. Note that Eq.~(\ref{eq:xi_extrapolation}) is
similar to $\xi = 0.306(1) - 0.805(2) \eta - 0.63(3) \eta^2$
in Ref.~\cite{Lee:2007} and gives $d\xi/d\eta|_{\eta=0} = -0.795(33)$
close to $-1.0(1)$ in Ref.~\cite{Lee:2006}.

\begin{figure}[htbp]
\begin{center}
\includegraphics[width=100mm]{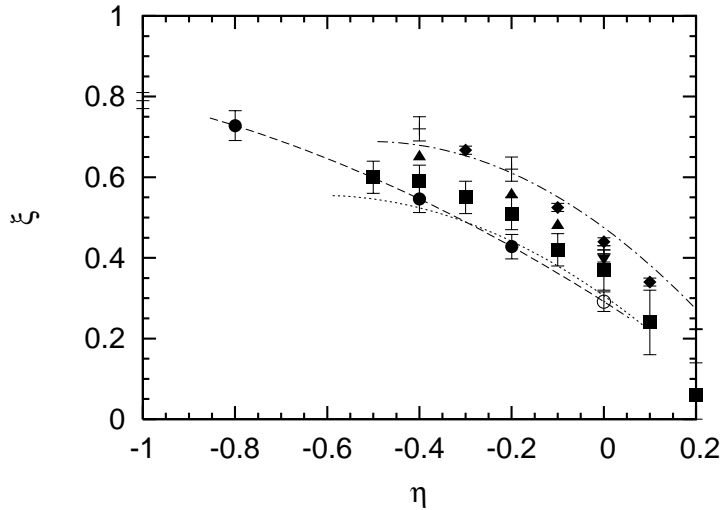}
\caption{$\eta \equiv 1/(k_F a_0)$ dependence of the energy
parameter $\xi$.  The solid circles and the dashed curve are our
Monte Carlo data and the best-fit,
Eq.~(\ref{eq:xi_extrapolation}), respectively.  Our $\xi$
extrapolated to the unitary limit is shown by the open circle. For
comparison, $\xi$ obtained by other works are also shown: the
dash-dotted curve is by the next-to-leading order $\epsilon$ expansion
\cite{Chen:2006wx}, the dotted curve is by the lattice Monte Carlo
calculation with the symmetric heavy-light ansatz \cite{Lee:2007},
the cross symbols are by the fixed-node diffusion Monte Carlo
calculation \cite{Astrakharchik:2004}, the solid up- and
down-triangles are by the fixed-node Green's function Monte Carlo
calculations at $\eta < 0$ \cite{Gezerlis:2008} and at $\eta = 0$
\cite{Carlson:2005kg}, respectively, and the solid squares are by
the determinantal quantum Monte Carlo calculation 
\cite{Bulgac:2008-1,Bulgac:2008-2}.} \label{Fig:xi_extrapolation}
\end{center}
\end{figure}

Figure~\ref{Fig:xi_extrapolation} shows the three $\xi$ values of
Eq.~(\ref{Eq:xi_finite_final}) and their best fit
Eq.~(\ref{eq:xi_extrapolation}), together with $\xi$ at the
unitary limit. In the figure, we also show several results of the
$\eta$ dependence of $\xi$ reported in the literature. They
include the two types of calculations: (i) analytical and (ii)
numerical.

(i) This type is an $\epsilon$-expansion calculation at the next-to-leading
order (NLO) \cite{Chen:2006wx} about four dimension (shown by the
dash-dotted curve). Note that a recent next-to-next-to-leading order
(NNLO) calculation at $\eta=0$ \cite{Arnold:2006} shows the
appearance of a problematic $\ln \epsilon$ contribution, but it is
argued to be infrared manageable.

(ii) The Monte Carlo types of calculation include the lattice Monte Carlo
calculation with the symmetric heavy-light ansatz \cite{Lee:2007}
(shown by the dotted curve), the diffusion Monte Carlo method
\cite{Astrakharchik:2004} (cross symbols), the fixed-node
Green's function Monte Carlo method at $\eta < 0$ (solid
up-triangles) \cite{Gezerlis:2008}, and at $\eta = 0$ (solid 
down-triangle) \cite{Carlson:2005kg}, and the determinantal
quantum Monte Carlo method (solid squares)
\cite{Bulgac:2008-1, Bulgac:2008-2}. Note that the previously
reported $\xi$ by the fixed-node Green's function Monte Carlo
calculations \cite{Carlson:2003,Carlson:2004} (solid
diamonds) were somewhat larger, being near that of 
Ref.~\cite{Astrakharchik:2004}.  We further note that no
thermodynamic limit is taken in these Monte Carlo calculations.

\begin{figure}[htbp]
\begin{center}
\includegraphics[width=100mm]{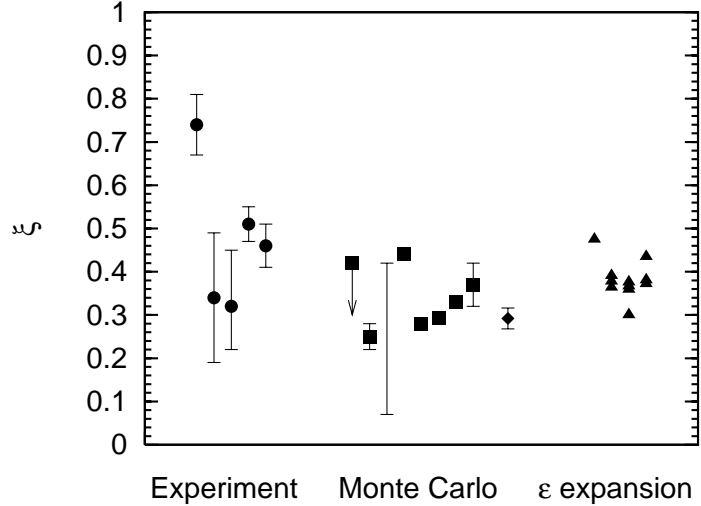}
\caption{$\xi$ at the unitary limit. The values of $\xi$
reported in the literature are shown in three groups from left to
right: those determined by atomic Fermi-gas
experiments (solid circles); by
various Monte Carlo calculations (solid squares); and
by $\epsilon$ expansions (triangles). 
The scale of the horizontal axis has no significance
but for separating data. $\xi$'s by the $\epsilon$-expansion
method on the right-hand side are divided into three subgroups as
explained in the text.  Our $\xi$ from
Eq.~(\ref{eq:xi_extrapolation}) is shown by the diamond at the
right-most location in the Monte Carlo group.  See the text for
the reference to each value of $\xi$. The figure is an
expanded version of a similar figure (Fig.~14) in
Ref.~\cite{Arnold:2006}.} \label{Fig:xi_comparison}
\end{center}
\end{figure}

In Fig.~\ref{Fig:xi_comparison}, we compare our result with
various values of the unitary-limit $\xi$ in the literature.
The figure is made by expanding a similar figure (Fig.~14) in
Ref.~\cite{Arnold:2006}. On the left-hand side of the figure, the
values of $\xi$ determined by atomic Fermi-gas experiments are
shown by solid circles: $\xi = 0.74(7)$ \cite{exp_Duke}, $\xi =
0.34(15)$ \cite{exp_ENS}, $\xi = 0.32^{+0.13}_{-0.10}$
\cite{exp_Innsbruck}, $\xi = 0.46(5)$ \cite{exp_Duke2}, and $\xi =
0.51(4)$ \cite{exp_Rice}.

In the middle of Fig.~\ref{Fig:xi_comparison}, the values of $\xi$
obtained by various Monte Carlo calculations are shown by the
solid squares: $\xi \le 0.42(1)$ \cite{Carlson:2005kg}, $\xi =
0.25(3)$ \cite{Lee:2006}, $0.07 \le \xi \le 0.42$
\cite{Lee:2005it}, $\xi \approx 0.44$ \cite{Bulgac:2005pj}, $\xi
\approx 0.28$ \cite{Lee:2007}, $\xi = 0.292(12)$ and $0.329(5)$
\cite{Lee:2008xs}, and $\xi = 0.37(5)$ \cite{Bulgac:2008-1,
Bulgac:2008-2}. For comparison, our result,
Eq.~(\ref{eq:xi_extrapolation}), is shown by the diamond.

On the right-hand side of the figure, the values of $\xi$ by the
$\epsilon$ expansions are shown by triangles in three groups: from
left to right, (1) the NLO $\epsilon$ expansion
\cite{Nishida:2006br}, (2) the Borel-Pad{\'e} approximation between
the NLO expansions about four and two dimensions
\cite{Nishida:2006eu}, (3) the Borel-Pad{\'e} approximation between
the NNLO expansion around four dimensions and the NLO expansion
about two dimensions \cite{Arnold:2006}, and (4) the Borel-Pad{\'e}
approximation between the NNLO expansions about four and two
dimensions \cite{Nishida:2008}.

Figure~\ref{Fig:xi_comparison} shows that our value of $\xi$,
0.292(24), is relatively small among the values shown.

\section{Pairing gap $\Delta$}

The pairing gap $\Delta$ at the unitary limit may be simply
related to $E_{g.s.}$ as $\sim 2 E_{g.s.}$ \cite{Carlson:2003}. 
To examine the relationship, we determined $\Delta$ for
$\eta \rightarrow 0$ by extrapolation.

\begin{figure}[htbp]
\begin{center}
\includegraphics[width=100mm]{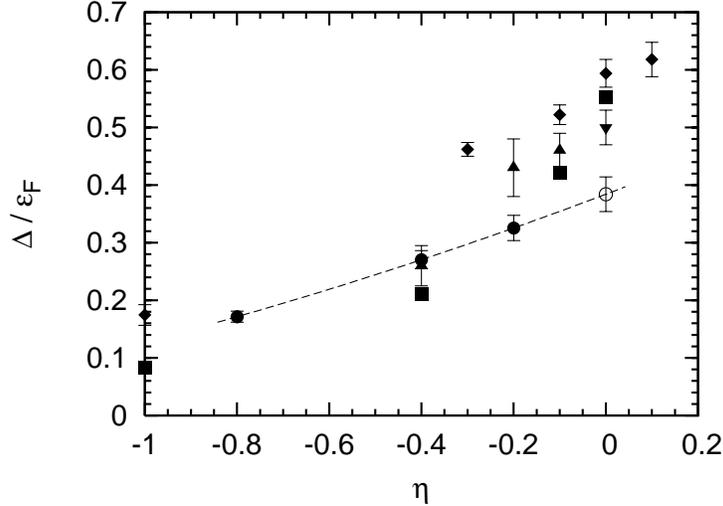}
\caption{Pairing gap $\Delta$ in the unit of
$\epsilon_F$ as a function of $\eta \equiv 1/(k_F a_0)$. The solid
circles are our Monte Carlo data, shown with statistical
uncertainties \cite{abeseki}. The dashed curve is the best fit to
the data by use of a quadratic function of $\eta$. $\Delta$ at the
unitary limit $\eta \rightarrow 0$ determined by extrapolation is
shown by an open circle. The solid squares, the solid
up-triangles, the solid down-triangle, and the solid diamonds are
the quantum Monte Carlo data by Bulgac 
{\it et al.}~\cite{Bulgac:2008-1,Bulgac:2008-2}, by Gezerlis 
{\it et al.}~\cite{Gezerlis:2008}, 
by Carlson {\it et al.}~\cite{Carlson:2005kg}, and
by Chang {\it et al.}~\cite{Carlson:2004}, respectively. Note that the
data by Chang {\it et al.}~at $\eta < 0$ are those quoted in
Refs.~\cite{Bulgac:2008-1,Bulgac:2008-2}.}
\label{Fig:Delta_unitary}
\end{center}
\end{figure}

In Fig.~\ref{Fig:Delta_unitary}, we show $\Delta$ in the
thermodynamic and continuum limits for the three values of
$\eta$ by taking from our previous work at the LO \cite{abeseki}.
The figure also shows the best-fit curve to the three $\eta$
values using the quadratic function
\begin{equation}
  \frac{\Delta}{\epsilon_F} = 0.384(30) + 0.303(27) \ \eta + 0.046(37) \ \eta^2,
\label{eq:Delta_extrapolation}
\end{equation}
from which $\Delta/\epsilon_F = 0.384(30)$ is determined by
setting $\eta = 0$.  The figure includes other quantum Monte Carlo
results: the solid squares are by Bulgac 
{\it et al.}~\cite{Bulgac:2008-1,Bulgac:2008-2}, the solid up-triangles are by
Gezerlis {\it et al.}~\cite{Gezerlis:2008}, the solid down-triangle is
by Carlson {\it et al.}~\cite{Carlson:2005kg}, and the solid diamonds
are by Chang {\it et al.}~\cite{Carlson:2004}.
Figure~\ref{Fig:Delta_unitary} is drawn similar to Fig.~1 of
Ref.~\cite{Bulgac:2008-1} and to Fig.~14 of
Ref.~\cite{Bulgac:2008-2}. Note that the data by Chang {\it et al.}~at
$\eta < 0$ are those quoted in
Refs.~\cite{Bulgac:2008-1,Bulgac:2008-2}.

The relation between $\Delta$ and $E_{g.s.}$ is found from that of
$\Delta$ and $\eta$ by eliminating $\epsilon_F$ from
Eqs.~(\ref{defxi}) and (\ref{eq:Delta_extrapolation}) and by using
$\xi$ of Eq.~(\ref{eq:xi_extrapolation}) at $\eta = 0$. Table II
summarizes our result. Our $\Delta/E_{g.s.}$ at the unitary limit is
2.19(35) and roughly confirms $\sim 2$ as suggested in 
Ref.~\cite{Carlson:2003}. Note that other quantum Monte Carlo
calculations in the literature also yield similar values:
$\Delta/E_{g.s.} = 2.5(3)$ \cite{Bulgac:2008-1, Bulgac:2008-2},
$2.3(1)$ \cite{Carlson:2004}, and $2.0(2)$ \cite{Carlson:2005kg}.

\begin{table}[htbp]
\caption{$\Delta/E_{g.s.}$ for various values of $\eta$}
\label{table:xi}
    \begin{center}
    \begin{tabular}{ccccc}
      \hline
      \hline
      $\eta$ & $\Delta / \epsilon_F$ & $\xi$ & $\Delta/E_{g.s.}$ \\
      \hline
      $-0.7997$ & $0.172(9)$ \ & $0.728(37)$ & $0.391(40)$ \\
      $-0.3999$ & $0.271(16)$ & $0.546(34)$ & $0.827(99)$ \\
      $-0.1999$ & $0.326(22)$ & $0.428(30)$ & $1.27(18)$ \ \\
      \ \, $0.0000$ & $0.384(30)$ & $0.292(24)$ & $2.19(35)$ \ \\
      \hline
      \hline
    \end{tabular}
  \end{center}
\end{table}

\section{Critical temperature $T_c$}
\label{subsec:comparison_Tc}

In the previous two sections, we examined the physical quantities
at zero temperature.  Thermodynamics at the unitary limit is of
much interest \cite{thermodynamics_unitary_limit, Bulgac:2008-1, Bulgac:2008-2}.
In this section, we examine a representative
thermodynamic quantity: the critical temperature of the phase
transition, $T_c$, at the unitary limit.

\begin{figure}[htbp]
\begin{center}
\includegraphics[width=100mm]{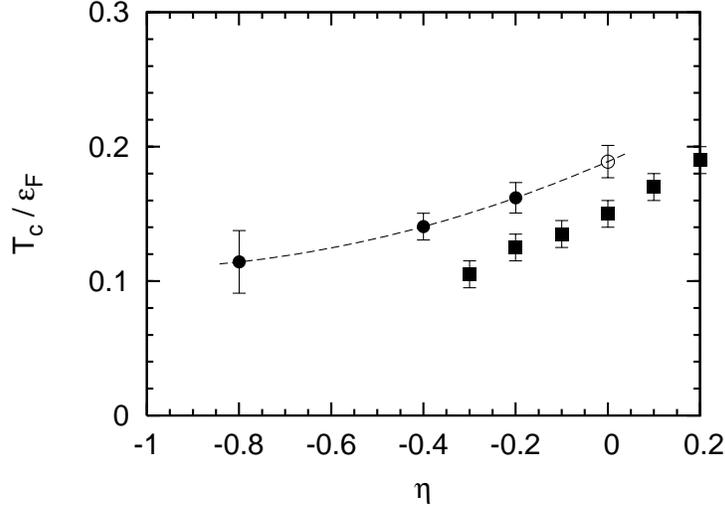}
\caption{$\eta$ dependence of the critical temperature $T_c$,
shown in the unit of $\epsilon_F$. Our Monte Carlo data at the
thermodynamic and continuum limits are shown by solid circles,
and the extrapolated unitary-limit point is shown by an open
circle. The extrapolation is made using the fit function,
Eq.~(\ref{eq:Tc_epsilonF}). For comparison, the solid squares show
a recent quantum Monte Carlo result \cite{Bulgac:2008-2}.}
\label{Fig:Tc_epsilonF}
\end{center}
\end{figure}

Figure~\ref{Fig:Tc_epsilonF} shows $T_c$ in the thermodynamic
and continuum limits for the three values of $\eta$ by taking from
our previous work at the LO \cite{abeseki}. The best fit to the
data is also shown in the figure by the dashed curve.  It is given
by
\begin{equation}
  \frac{T_c}{\epsilon_F} = 0.189(12) + 0.149(22) \ \eta + 0.069(36) \ \eta^2,
\label{eq:Tc_epsilonF}
\end{equation}
which yields the extrapolated value $T_c/\epsilon_F = 0.189(12)$
at $\eta = 0$, or at the unitary limit. For comparison, in the
figure we show $T_c/\epsilon_F$ by a recent quantum Monte Carlo
calculation \cite{Bulgac:2008-2}.  While our values are somewhat
larger than those of Ref.~\cite{Bulgac:2008-2}, both show similar 
$\eta$ dependence on $T_c$. This
$\eta$-$T_c$ dependence is also observed by an
$\epsilon$-expansion calculation \cite{Nishida:2006rp}.  Our $T_c$
is tabulated for various values of $\eta$ in Table III.

\begin{table}[htbp]
\caption{$T_c/\epsilon_F$ for various values of $\eta$.}
\label{table:Tc}
    \begin{center}
    \begin{tabular}{cc}
      \hline
      \hline
      $\eta$ & $T_c / \epsilon_F$ \\
      \hline
      $-0.7997$ & $0.114(23)$ \\
      $-0.3999$ & $0.141(10)$ \\
      $-0.1999$ & $0.162(11)$ \\
      \ \, $0.0000$ & $0.189(12)$ \\
      \hline
      \hline
    \end{tabular}
  \end{center}
\end{table}

\begin{figure}[htbp]
\begin{center}
\includegraphics[width=100mm]{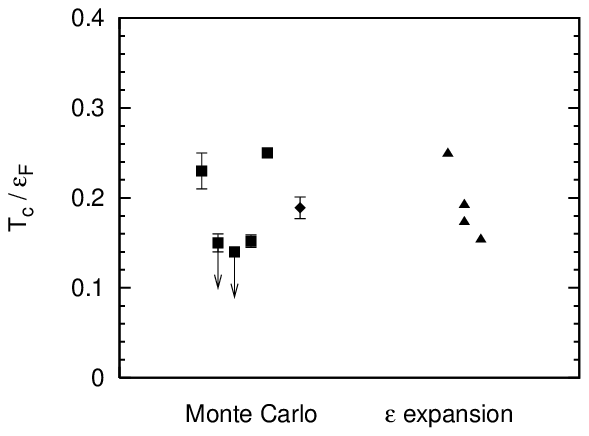}
\caption{Critical temperature $T_c$ at the unitary limit
appearing in the literature and our extrapolated $T_c$ (all shown
in the unit of $\epsilon_F$). The solid squares (with statistical
uncertainties) on the left-hand side of the figure are by Monte
Carlo calculations \cite{Bulgac:2005pj, Bulgac:2007,
Bulgac:2008-2, Lee:2005it, burovski, Akkineni-Tc}, and the solid
triangles on the right-hand side are by $\epsilon$ expansion
calculations \cite{Nishida:2006rp}. Our $T_c$ is shown by the
solid diamond with the statistical uncertainty. The horizontal
scale is used merely for the separation of each $T_c$ value.  The
$T_c$'s by the $\epsilon$ expansion calculations are grouped into
two in the same way as in Fig.~\ref{Fig:xi_comparison}.  See the
text for the reference of each $T_c$.}
\label{Fig:Tc_epsilonF_comparison}
\end{center}
\end{figure}

In Fig.~\ref{Fig:Tc_epsilonF_comparison}, we summarize various
values of $T_c$ at the unitary limit that are reported in the
literature, together with ours for comparison.  The figure
includes the two groups of the $T_c$ determination: by Monte Carlo
calculations and by the $\epsilon$-expansion method.

The Monte Carlo results include $T_c/\epsilon_F = 0.23(2)$ by
the determinantal quantum Monte Carlo method on $N_s = 6^3$ and
$8^3$ lattices \cite{Bulgac:2005pj, Bulgac:2007}; $T_c/\epsilon_F
< 0.15(1)$ by the determinantal quantum Monte Carlo method on
$N_s = 6^3$, $8^3$, and $10^3$ lattices \cite{Bulgac:2008-2};
$T_c/\epsilon_F < 0.14$ by the hybrid Monte Carlo method on
$N_s = 4^3$, $5^3$, and $6^3$ lattices \cite{Lee:2005it};
$T_c/\epsilon_F = 0.152(7)$ by a diagrammatic determinantal
quantum Monte Carlo method with the finite-size scaling
\cite{burovski}; and $T_c/\epsilon_F \approx 0.25$ by a
restricted path integral Monte Carlo method \cite{Akkineni-Tc}.

The $\epsilon$-expansion results \cite{Nishida:2006rp} include
$T_c / \epsilon_F \approx 0.249$ and $T_c/\epsilon_F \approx
0.153$ by up to the NLO in the $\epsilon$ expansion about the
four and two dimensions, respectively; and $T_c/\epsilon_F =
0.173$ and $0.192$ by the Borel-Pad{\'e} approximation
between the four and two dimensions. As noted in the case of the
$\xi$ determination, the $\epsilon$ expansion seems to
indicate a possibly problematic behavior at higher orders
\cite{Arnold:2006}.

Our $T_c$ at the thermodynamic and continuum limit is also shown
in the Monte Carlo group of Fig.~\ref{Fig:Tc_epsilonF_comparison}.
Our result $T_c/\epsilon_F \approx 0.189(12)$ seems to be
consistent with others, especially with that of Ref.~\cite{burovski},
which is the only other $T_c$ obtained in the thermodynamic
and continuum limits.

\section{Conclusion}

We extrapolate our Monte Carlo results for low-density neutron
matter \cite{abeseki} to the unitary limit for the following
quantities: the ground-state energy $E_{g.s.}$, the pairing gap
$\Delta$ at $T \approx 0$, and the critical temperature of the
normal-to-superfluid phase transition $T_c$.  All quantities show
a smooth extrapolation to the limit in terms of $\eta =1/(k_F
a_0)$. Although no accurate determination of these quantities is
yet available, our extrapolated values are in reasonable agreement
with those in the literature.  Our successful extrapolation
suggests that much of the physics of low-density neutron matter
[of about $(10^{-4}-10^{-2})\rho_0$] is similar to the physics
of the attractively interacting fermion system at the unitary
limit.

\bigskip
\bigskip
\centerline{\bf ACKNOWLEDGMENTS}
\bigskip

We thank M.~Ueda for suggesting comparison of our neutron matter
calculation to those of the unitary limit and D. Lee for his
useful comments after reading the manuscript. 
We thank R.~McKeown for the
generous hospitality at Kellogg Radiation Laboratory at Caltech,
where a part of the project was carried out.
Some of this work was also performed at the Yukawa Institute for 
Theoretical Physics (YITP), Kyoto University. 
R.~S.~thanks the YITP for its warm hospitality. 
Lattice calculations were carried
out on the Seaborg, the Bassi, and the Franklin at the National
Energy Research Scientific Computing Center, which is supported by
the Office of Science of the U.S.~Department of Energy under
Contract No.~DE-AC03-76SF00098, and at Titech Grid, and TSUBAME,
Tokyo Institute of Technology, Japan. 
This work is being supported by the U.S.~Department of Energy under Grant
No.~DE-FG02-87ER40347 at CSUN.

\vfill\eject



\begin{thebibliography}{99}

\bibitem{abeseki}
  T.~Abe and R.~Seki, Phys. \ Rev. \ C {\bf 79}, 054002 (2009).

\bibitem{eft}
  R.~Seki, U.~van Kolck, and M.~J.~Savage, 
  {\it Nuclear Physics with Effective Field Theory;  Proceedings of the Joint
  Caltech/INT Workshop} (World Scientific, Singapore, 1998);
  P.~F.~Bedaque, M.~J.~Savage, R.~Seki, and U.~van Kolck, 
  {\it Nuclear Physics with Effective Field Theory II; An INT
  Workshop} (World Scientific, Singapore, 2000).

\bibitem{crossover}
  A.~J.~Leggett, J. \ Phys. \ (Paris) {\bf 41}, C7--19 (1980);
  P.~Nozieres and S.~Schmitt-Rink, J. \ Low Temp. \ Phys. \ {\bf 59},
  195 (1985);
  Q.~Chen, J.~Stajic, S.~Tan, and K.~Levin, Phys. \ Rep. \ {\bf 412}, 1 (2005)
  and references therein.

\bibitem{abook}
  D.~M.~Brink and R.~A.~Broglia,
  {\it Nuclear Superfluidity; Pairing in Finite Systems}
  (Cambridge University, Cambridge, England, 2005),
  and references therein.

\bibitem{Randeria}
  M.~Randeria, in {\it Bose-Einstein Condensation}, edited by
  A.~Griffin, D.~Snoke, and S.~Stringari (Cambidge University,
  Cambridge, England, 1994) p.~355.

\bibitem{burovski}
  E.~Burovski, N.~Prokof'ev, B.~Svistunov, and M.~Troyer, Phys. \ Rev. \
  Lett. \ {\bf 96}, 160402 (2006); New J. \ Phys. \ {\bf 8}, 153 (2006), 
  and references therein.

\bibitem{thermodynamics_unitary_limit}
  H.~Heiselberg, Phys. \ Rev. \ A {\bf 63}, 043606 (2001);
  J.~R.~Engelbrecht, M.~Randeria, and C.~A.~R.~S{\'a} de Melo,
  Phys. \ Rev. \ B {\bf 55}, 15153 (1997);
  C.~A.~R.~S{\'a} de Melo, M.~Randeria, and J.~R.~Engelbrecht,
  Phys. \ Rev. \ Lett. \ {\bf 71}, 3202 (1993).

\bibitem{Noyes:1972}
  H.~P.~Noyes, Annu. \ Rev. \ Nucl. \ Sci. \ {\bf 22}, 465 (1972).

\bibitem{ask}
  T.~Abe, R.~Seki and A.~N.~Kocharian,
  Phys.\ Rev.\ C {\bf 70}, 014315 (2004); 
  {\bf 71}, 059902(E) (2005).

\bibitem{Lee:2006}
  D.~Lee, Phys.\ Rev.\ B {\bf 73}, 115112 (2006).

\bibitem{Lee:2005it}
  D.~Lee and T.~Sch{\"a}fer, Phys.\ Rev.\ C {\bf 73}, 015202 (2006).

\bibitem{Lee:2008xs}
  D.~Lee, Phys. \ Rev. \ C {\bf 78}, 024001 (2008).

\bibitem{Lee:2008fa}
  D.~Lee, arXiv:0804.3501.

\bibitem{ck}
  J.-W.~Chen and D.~B.~Kaplan,
  Phys. \ Rev. \ Lett. \ {\bf 92}, 257002 (2004).

\bibitem{bira}
  U.~van~Kolck, Nucl. \ Phys. \ {\bf A645}, 273 (1999).

\bibitem{lepage}
  G.~P.~Lepage, Lecture at the VIII Jorge Andr\'{e} Swieca Summer School, 
  Brazil, 1997; arXiv:nucl-th/9706029.

\bibitem{sv}
  R.~Seki and U.~van Kolck, Phys.\ Rev.\ C {\bf 73}, 044006 (2006).

\bibitem{watson}
  G.~N.~Watson, Q. \ J. \ Math. \ (Oxford) {\bf 10}, 266 (1939);
  A.~A.~Maradudin, E.~W.~Montroll, G.~H.~Weiss, R.~Herman, and H.~W.~Milnes,
  Acad. \ R. \ Belg. \ Cl. \ Sci. \ Mem. \ Coll. \
  $4^o$ (2) 14 (1960) No. \ 7.

\bibitem{LohGubernatis1992}
  E.~Y.~Loh Jr. and J.~E.~Gubernatis, in {\it Electronic Phase
  Transitions}, edited by W.~Hanke and Yu.~V.~Kopaev (Elsevier,
  Amsterdam, 1992).

\bibitem{dosSantos2003}
  R.~R.~dos Santos, Braz. \ J. \ Phys. \ {\bf 33}, 36 (2003).

\bibitem{LatticeQCD}
  H.~J.~ Rothe, {\it Lattice Gauge Theories: An Introduction,
  World Scientific Lecture Notes in Physics}, 3rd ed. (World
  Scientific Pub. Co., 2005);
  J. Smit, {\it Introduction to Quantum Fields on a Lattice}
  (Cambridge Univ. Press, 2002);
  I. Montvay and G. M\"{u}nster,
  {\it Quantum Fields on a Lattice, Cambridge Monographs on
  Mathematical Physics} (Cambridge Univ. Press, 1997);
  M. Creutz, {\it Quarks, Gluons and Lattices,
  Cambridge Monographs on Mathematical Physics}
  (Cambridge Univ. Press, 1985);
  and references quoted therein.

\bibitem{Lee:2007}
  D.~Lee,
  Eur. \ Phys. \ J. \ A {\bf 35}, 171 (2008).

\bibitem{Chen:2006wx}
  J.~W.~Chen and E.~Nakano,
  Phys. \ Rev. \ A {\bf 75}, 043620 (2007).

\bibitem{Arnold:2006}
  P.~Arnold, J.~E.~Drut, D.~T.~Son,
  Phys. \ Rev. \ A {\bf 75}, 043605 (2007).

\bibitem{Astrakharchik:2004}
  G.~E.~Astrakharchik, J.~Boronat, J.~Casulleras, and S.~Giorgini,
  Phys. \ Rev. \ Lett. \ {\bf 93}, 200404 (2004).

\bibitem{Gezerlis:2008}
  A.~Gezerlis, and J.~Carlson,
  Phys. \ Rev. \ C {\bf 77}, 032801(R) (2008).

\bibitem{Carlson:2005kg}
  J.~Carlson and S.~Reddy,
  Phys. \ Rev. \ Lett. \ {\bf 95}, 060401 (2005).

\bibitem{Bulgac:2008-1}
  A.~Bulgac, J.~E.~Drut, P.~Magierski, and G.~Wlazlwski,
  arXiv:0801.1504.

\bibitem{Bulgac:2008-2}
  A.~Bulgac, J.~E.~Drut, and P.~Magierski, 
  Phys. \ Rev. \ A {\bf 78}, 023625 (2008).

\bibitem{Carlson:2003}
  J.~Carlson, S.~Y.~Chang, V.~R.~Pandharipande, and K.~E.~Schmidt, 
  Phys. \ Rev. \ Lett. \ {\bf 91}, 050401 (2003).

\bibitem{Carlson:2004}
  S.~Y.~Chang, V.~R.~Pandharipande, J.~Carlson, and K.~E.~Schmidt,
  Phys. \ Rev. \ A {\bf 70}, 043602 (2004).

\bibitem{exp_Duke}
  K.~M.~O'Hara, S.~L.~Hemmer, M.~E.~Gehm, S.~R.~Granade, and
  J.~E.~Thomas, Science {\bf 298}, 2179 (2002); M.~E.~Gehm, S.~L.~Hemmer,
  S.~R.~Granade, K.~M.~O'Hara, and J.~E.~Thomas,
  Phys. \ Rev. \ A {\bf 68}, 011401(R) (2003).

\bibitem{exp_ENS}
  T.~Bourdel, L.~Khaykovich, J.~Cubizolles, J.~Zhang, F.~Chevy,
  M.~Teichmann, L.~Tarruell, S.~J.~J.~M.~F.~Kokkelmans, and C.~Salomon,
  Phys. \ Rev. \ Lett. \ {\bf 93}, 050401 (2004).

\bibitem{exp_Innsbruck}
  M.~Bartenstein, A.~Altmeyer, S.~Riedl, S.~Jochim, C.~Chin,
  J.~H.~Denschlag, and R.~Grimm,
  Phys. \ Rev. \ Lett. \ {\bf 92}, 120401 (2004).

\bibitem{exp_Duke2}
  J.~Kinast, A.~Turlapov, J.~E.~Thomas, Q.~Chen, J.~Stajic, and
  K.~Levin, Science {\bf 307}, 1296 (2005).

\bibitem{exp_Rice}
  G.~B.~Partridge, W.~Li, R.~I.~Kamar, Y.~Liao, and R.~G.~Hulet,
  Science {\bf 311}, 503 (2006).

\bibitem{Bulgac:2005pj}
  A.~Bulgac, J.~E.~Drut, and P.~Magierski,
  Phys. \ Rev. \ Lett. \ {\bf 96}, 090404 (2006).

\bibitem{Nishida:2006br}
  Y.~Nishida and D.~T.~Son,
  Phys. \ Rev. \ Lett. \ {\bf 97}, 050403 (2006).

\bibitem{Nishida:2006eu}
  Y.~Nishida and D.~T.~Son,
  Phys. \ Rev. \ A {\bf 75}, 063617 (2007).

\bibitem{Nishida:2008}
  Y.~Nishida, 
  Phys. \ Rev. \ A {\bf 79}, 013627 (2009).

\bibitem{Nishida:2006rp}
  Y.~Nishida,
  Phys. \ Rev. \ A {\bf 75}, 063618 (2007).

\bibitem{Bulgac:2007}
  A.~Bulgac, J.~E.~Drut, and P.~Magierski,
  Phys. \ Rev. \ Lett. \ {\bf 99}, 120401 (2007).

\bibitem{Akkineni-Tc}
  V.~K.~Akkineni, D.~M.~Ceperley, and N.~Trivedi,
  arXiv:cond-mat/0608154.


\end{thebibliography}
\end{document}